\documentclass[12pt]{article}
\newcommand{\be}{\begin{equation}}
\newcommand{\ee}{\end{equation}}
\def\n{\noindent}
\catcode `\@=11
\catcode `\@=12
 
\begin{document}
\title{PLANE SYMMETRIC DOMAIN WALL IN LYRA GEOMETRY}
\author{Anirudh Pradhan\thanks{Corresponding Author e-mail: acpradhan@yahoo.com, 
apradhan@mri.ernet.in}, Aotemshi I. \\
Department of Mathematics,\\ Hindu Post-graduate College\\
Zamania-232 331, Ghazipur, U. P., India  \\ 
and \\ G. P. Singh \\   Department of Mathematics\\
Visvesvaraya Regional College of Engineering,\\ Nagpur-440 011, India.\\ 
\hspace{1cm} e-mail: gps@vrce.ernet.in}
\date{}
\maketitle
\begin{abstract} 
In this paper general solutions are found for domain walls in Lyra geometry
in the plane symmetric spacetime metric given by Taub. Expressions for the 
energy density and pressure of domain walls are derived in both cases of 
uniform and time varying displacement field $\beta$. It is also shown that 
the results obtained by Rahaman et al [IJMPD, {\bf 10}, 735 (2001)] are 
particular case of our solutions. Finally, the geodesic equations and 
acceleration of the test particle are discussed. 
\end{abstract}
\smallskip
\n KEYWORDS : Domain Wall, Cosmology, Lyra Geometry.\\
\n PACS No. : 98.80.H, 04.20.Jb
\newpage
\section{Introduction}
  \vspace*{-0.5pt}
\noindent
 The study of topological defects in cosmology relevant to structure formation 
and evolution of the universe has been widely studied \cite{ref1}. Of all the 
topological defects domain walls are deceptively simple to study. It is 
established that absolutely stable domain walls that have a mass per unit area 
greater than $4\times 10^{-3}g/cm^{2} [(10 Me V)^3$ in units where $h = c = 1$] 
are cosmological disaster \cite{ref2,ref3}. However, unstable 
(but perhaps long-lived) domain walls could have been present in the early 
universe \cite{ref4}. Furthermore, light domain walls might even be present 
today \cite{ref5}. In either case domain walls may play an important role 
in the universe.
\newline
\par

Domain walls form when a discrete symmetry is spontaneously broken \cite{ref2}. In  
simplest models, symmetry breaking is accomplished by a real scalar field $\phi$
whose vacuum manifold is disconnected. [For example, suppose that the scalar 
potential for $\phi$ is $U(\phi) = \lambda (\phi^{2} - \mu^{2})^2$. The vacuum 
manifold for $\phi$ then consists of the two points [$\phi = \mu $ and 
$\phi = - \mu$]. After symmetry breaking, different regions of the universe can 
settle into different parts of the vacuum with domain walls forming the boundaries 
between these regions. The stress energy for a static, plane-symmetric domain wall 
consists of a positive surface energy density and a surface tension equal in 
magnitude to the surface energy \cite{ref3}. We note, however that this analysis 
neglects the effects of gravity \cite{ref6}. Locally, the stress energy for a wall 
of arbitrary shape is similar to that of a plane-symmetric wall having both surface 
energy density and surface tension. Closed-surface domain walls collapse due to 
their surface tension. However, the details of the collapse for a wall with 
arbitrary shape and finite thickness are largely unknown.
\newline
\par

The spacetime of cosmological domain walls has now been a subject of interest
for more than a decade since the work of Vilenkin and Ipser and Sikivie \cite{ref7,ref8}
who use Israel's thin wall formalism \cite{ref9} to compute the gravitational field of
an infinitesimally thin planar domain wall. After the original work by Vilenkin,
Ipser and Sikivie \cite{ref7,ref8} for thin walls, attempts focused on trying to find a 
perturbative expansion in the wall thickness \cite{ref10,ref6}. With the proposition by 
Hill, Schramn and Fry \cite{ref11} of a late phase transition with thick domain walls, 
there was some effort in finding exact thick solution \cite{ref12,ref13}. Recently, 
Bonjour et al \cite{ref14} considered gravitating thick domain wall solutions with 
planar and reflection symmetry in the Goldstone model. Bonjour et al \cite{ref15} 
also investigated the spacetime of a thick gravitational domain wall for a general 
potential $V(\phi)$. Jensen and Soleng \cite{ref16} have studied anisotropic domain 
walls where the solution has naked singularities and the generic solution is unstable 
to Hawking decay.
\newline
\par

The universe is spherically symmetric and the matter distribution in
it is on the whole isotropic and homogeneous. But during the early stages of 
evolution, it is unlikely that it could have had such a smoothed out picture. 
Hence we consider plane symmetry which provides an opportunity for the study 
of inhomogeneity. Most studies in cosmology involve a perfect fluid. Large entropy 
per baryon and the remarkable degree of isotropy of the cosmic microwave background 
radiation, suggest that we should analyse dissipative effects in cosmology. Further,
there are several processes which are expected to give rise to viscous effect. 
These are the decoupling of neutrinos during the radiation era and the recombination 
era \cite{ref17}, decay of massive super string modes into massless modes \cite{ref18},
gravitational string production \cite{ref19,ref20} and particle creation effect in 
grand unification era \cite{ref21}. It is known that the introduction of bulk 
viscosity can avoid the big bang singularity. Thus, we should consider the presence 
of a material distribution other than a perfect fluid to have realistic 
cosmological models (see Gr\o n \cite{ref22} for a review on cosmological 
models with bulk viscosity). 
\newline
\par

To study thick domain walls, one can study the field equations as well as equations 
of the domain walls treated as the self interacting scalar field. A thick domain wall 
can be viewed as a soliton-like solution of the scalar field equation coupled with 
gravity. In order to determine the gravitational field one has to solve Einstein's 
equation with an energy momentum tensor $T_{\mu \nu}$ describing a scalar field 
$\phi$ with self-interactions contained in a potential 
$V(\phi)$ \cite{ref6,ref7,ref8,ref12}.
\begin{equation}
\label{eq1}
 T_{\mu \nu} = \delta_{\mu} \phi \delta_{\nu} \phi - g_{\mu \nu} \left( \frac{1}{2}
\delta_{\sigma}\phi \delta^{\sigma}\phi - V(\phi)\right)   
\end{equation}
\noindent 
In another approach to study the phenomenon one has to assume the energy momentum 
tensor in the form
\begin{equation}
\label{eq2}
T_{i k} = \rho ( g_{ik} + W_{i}W_{k}) + p W_{i} W_{k}, ~ ~ W_{i} W^{K} = - 1  
\end{equation}
\noindent 
where $\rho$ is the energy density of the wall, $p$ is the pressure in the 
direction normal to the plane of the wall and $W_{i}$ is a unit space like vector 
in the same direction \cite{ref23}.
\newline
\par
In 1951 Lyra \cite{ref24} proposed a modification of Riemannian geometry
by introducing a gauge function into the structureless manifold,
as a result of which the cosmological constant arises naturally from
geometry. This bears a remarkable resemblance to Weyl's geometry \cite{ref25}. 
But in Lyra's geometry, unlike Weyl's, the connection is metric preserving as in
Riemannian; in other words, length transfers are integrable. Lyra also 
introduced the notion of a gauge and in the ``normal'' gauge the curvature 
scalar is identical to that of Weyl. In subsequent investigations Sen \cite{ref26}, 
Sen and Dunn \cite{ref27}  proposed a new scalar-tensor theory of gravitation
and constructed an analog of the Einstein field equations based 
on Lyra's geometry. It is, thus, possible \cite{ref26} to construct a geometrised 
theory of gravitation and electromagnetism along the lines of Weyl's ``unified''
field theory without inconvenience of non-integrability length
transfer. Halford \cite{ref28} has pointed out that the constant vector displacement
field $\phi_i$ in Lyra's geometry plays the role of cosmological
constant $\Lambda$ in the normal general relativistic treatment. It
is shown by Halford \cite{ref29} that the scalar-tensor treatment based on
Lyra's geometry predicts the same effects, within observational limits
as the Einstein's theory. Several investigators viz. Sen and Vanstone \cite{ref30}, 
Bhamra \cite{ref31}, Karade and Borikar \cite{ref32}, Kalyanshetti and Wagmode 
\cite{ref33}, Reddy and Innaiah \cite{ref34}, Beesham \cite{ref35}, Reddy and 
Venkateswarlu \cite{ref36}, Soleng \cite{ref37}, Singh and Singh \cite{ref38}, 
Singh and Desikan \cite{ref39}, Pradhan and Vishwakarma \cite{ref40}, 
Pradhan et al \cite{ref41,ref42} have studied cosmological models based on Lyra's 
manifold  with a constant displacement field vector. However, this restriction of 
the displacement field to be constant is merely one of convenience and there is 
no {\it a priori} reason for it. Soleng \cite{ref37} has pointed out that the 
cosmologies based on Lyra's manifold with constant gauge vector $\phi$ will either 
include a creation  field and be equal to Hoyle's creation field cosmology 
{\cite{ref43}$-$\cite{ref45}} or contain a special vacuum field which together with 
the gauge vector term may be considered as a cosmological term. In the latter case 
the solutions are equal to the general relativistic cosmologies with a cosmological term.\\
\newline
\par

In this paper we obtain all possible general solutions for domain wall in 
Lyra geometry in the plane symmetric spacetime metric given by Taub. Expressions 
for the energy density and pressure of domain walls are obtained in both cases of 
uniform and time varying displacement field $\beta$. We have also shown that the 
result obtained by Rahaman et al \cite{ref23} is a special case of our solutions.
\newline
\par
\section{Field Equations}
\noindent
In this section we shall consider the field equations, in normal gauge for Lyra's
manifold, obtained by Sen \cite{ref26} as
\begin{equation}
\label{eq3}
R_{ij} - \frac{1}{2} g_{ij} R + \frac{3}{2} \phi_i \phi_j
- \frac{3}{4} g_{ij} \phi_k \phi^k = - 8 \pi G T_{ij} 
\end{equation}
\noindent 
for study domain walls. The energy momentum tensor $T_{ij}$ in comoving 
coordinates for thick domain walls take the form
\begin{equation}
\label{eq4}
T^{0}_{0} = T^{2}_{2} = T^{3}_{3}= \rho,~ ~  T^{1}_{1} = - p_{1},~ ~  T^{0}_{1} = 0
\end{equation}
\noindent 
and displacement vector $\phi_{i}$ is defined by $\phi_{i} = (0, 0, 0, \beta)$,
where $\beta$ may be considered constant as well as function of time coordinate 
like cosmological constant in Einstein's theory of gravitation.\\
\noindent
We consider the most general plane symmetric spacetime metric suggested 
by Taub \cite{ref46} 
\begin{equation}
\label{eq5}
ds^2 = e^A (dt^2 - dz^2) - e^B (dx^2 + dy^2)
\end{equation}
\noindent 
where $A$ and $B$ are functions of $t$ and $z$.\\
\noindent 
Using equation (\ref{eq4}) the field equations (\ref{eq3}) for the metric (5) reduce to
\begin{equation}
\label{eq6}
\frac{e^{-A}}{4}(-4B'' - 3B'^2 + 2A' B') + \frac{e^{-A}}{4} (\dot B^2 + 2 \dot B\dot A)
- \frac{3}{4} e^{-A} \beta^2 = 8 \pi \rho
\end{equation}
\begin{equation}
\label{eq7}
\frac{e^{-A}}{4}(- B'^2 - 2 B' A') + \frac{e^{-A}}{4}(- 4 \ddot B + 3 \dot B^2 - 2 \dot A 
\dot B) + \frac{3}{4} e^{-A} \beta^2 = - 8 \pi p_{1}
\end{equation}
\begin{equation}
\label{eq8}
\frac{e^{-A}}{4}[-2(A'' + B'') - B'^2] + \frac{e^{-A}}{4}[2(\ddot A + \ddot B) + \dot B^2] 
+ \frac{3}{4} e^{-A} \beta^2 = 8 \pi \rho
\end{equation}
\begin{equation}
\label{eq9}
- \dot {B}' + \dot B(A' - B') + \dot A B' = 0
\end{equation}
\noindent
 In order to solve the above set of field equations we assume the separable form of
the metric coefficients as follows
\begin{equation}
\label{eq10}
A = A_1(z) + A_2(t),~ ~ ~ B = B_1(z) + B_2(t)
\end{equation}
\noindent 
From Eqs. (\ref{eq9}) and (\ref{eq10}), we obtain
\begin{equation}
\label{eq11}
\frac{A'_{1}}{B'_{1}} = \frac{(\dot B_2 -\dot A_2)}{B_2} = m,
\end{equation}
\noindent
where $m$ is considered as separation constant
 
\noindent
Eq. (\ref{eq11}) yields the solution
\begin{equation}
\label{eq12}
A_1 = m B_1
\end{equation}
\begin{equation}
\label{eq13}
A_2 = (1 - m) B_2
\end{equation}
\noindent 
Again, subtracting Eq. (\ref{eq8}) from Eq. (\ref{eq6}) and using Eq. (\ref{eq10}), 
we obtain
\begin{equation}
\label{eq14}
A''_{1} - B''_{1} - B'^2_{1} + A'_{1} B'_{1} = \ddot A_{2} + \ddot B_{2} - \dot A_{2} 
\dot B_{2} + 3 \beta^2 = k,
\end{equation}
\noindent
where $k$ is another separation constant.
\noindent 
With the help of Eqs (\ref{eq12}) and (\ref{eq13}), Eq. (\ref{eq14}) may be written as
\begin{equation}
\label{eq15}
(m - 1)[B''_{1} + B'^2_{1}] = k
\end{equation}
\begin{equation}
\label{eq16}
(2 - m) \ddot B_{2} + (m - 1) \dot B^2_{2} = k - 3 \beta^2
\end{equation}
\section{Solutions of the field equations}
    In this section we shall obtain exact solutions for thick domain walls in different
cases.\\
\noindent
 Using the substitution $u = e^{B_1}$ and $a = \frac{k}{1- m}$, Eq. (\ref{eq15}) 
takes the form
\begin{equation}
\label{eq17}
u'' + au = 0
\end{equation}
\noindent
which has solution
\begin{equation}
\label{eq18}
e^{B_1} =  u = \left[ \begin{array}{ll} c_1 sinh(z\sqrt{\mid a \mid}) + 
c_2 cosh(z\sqrt{\mid a \mid})  & \mbox { when $a<0$}\\ 
c_1 + c_2 z & \mbox { when $a = 0$} \\
c_1 sin(z\sqrt{a}) + c_2 cos(z\sqrt{a})  & \mbox { when $a>0$} \end{array} \right. 
\end{equation}
\noindent
where $c_1$ and $c_2$ are integrating constants. Eq. (\ref{eq18}) represent the 
complete solution of the differential Eq. (\ref{eq17}). It may be noted that 
Rahaman et al \cite{ref23} have obtained a particular solution for the case 
$a < 0$. Their solution can be obtained from Eq. (\ref{eq18}) by taking $c_1 = 0$ 
and $c_2 = 2$ for the case $a<0$.\\
\noindent
Eq. (\ref{eq16}) may be written as
\begin{equation}
\label{eq19}
\ddot B_2 - \frac{( 1- m)}{(2 - m)}\dot B^2_{2} + \frac{3}{(2 - m)} \beta^2 = 
\frac{k}{2 - m}
\end{equation}
\noindent
Now we shall consider uniform and time varying displacement field $\beta$ separately.\\
\noindent
{\bf Case I :} Uniform displacement field ($\beta = \beta_0$, constant) \\
\noindent
By use of the transformation $v = e^{-\frac{(1 - m)}{(2 - m)}B_2}$, equation (\ref{eq19}) 
reduces to
\begin{equation}
\label{eq20}
\ddot v + b v = 0,
\end{equation}
\noindent
where $b = \frac{(1 - m) (k - 3\beta^2_0)}{(2 - m)^2}$.
\noindent
 Again, it can be easily seen that Eq. (\ref{eq20}) possesses the solution
\begin{equation}
\label{eq21}
e^{-\frac{(1 - m)}{(2 -m)} B_2} = v = \left[ \begin{array}{ll}
            \bar{c}_1 sinh(t\sqrt{\mid b \mid}) + \bar{c}_2 cosh(t\sqrt{\mid b \mid})  
& \mbox { when $b<0$}\\
\bar{c}_1 + \bar{c}_2 t                                         & \mbox { when $b=0$} \\
            \bar{c}_1 sin(t\sqrt{b}) + \bar{c}_2 cos(t\sqrt{b})  & \mbox { when $b>0$}
            \end{array} \right. 
\end{equation}
\noindent
where $\bar{c}_1$ and $\bar{c}_2$ are integrating constants. Hence the metric coefficients 
have the explicit forms as
\begin{equation}
\label{eq22}
e^A = \left[ \begin{array}{ll}
            \{c_1 sinh(z\sqrt{\mid a \mid}) + c_2 cosh(z\sqrt{\mid a \mid})\}^{m}\times &\\
            \{\bar{c}_1 sinh(t\sqrt{\mid b \mid}) + \bar{c}_2 cosh(t\sqrt{\mid b \mid})\}^
            {(m - 2)}   & \mbox { when $a<0,~~ b<0$}\\
            \{c_1 + c_2 z\}^{m} \{ \bar{c}_1 + \bar{c}_2 t\}^{(m - 2)}     
            & \mbox { when $a = 0,~~ b = 0$} \\
            \{c_1 sin(z\sqrt{a}) + c_2 cos(z\sqrt{a})\}^{m}\times &\\
            \{\bar{c}_1 sin(t\sqrt{b}) + \bar{c}_2 cos(t\sqrt{b})\}^{(m - 2)}  
           & \mbox {when $a>0,~~ b>0$}
            \end{array} \right. 
\end{equation}
\begin{equation}
\label{eq23}
e^B = \left[ \begin{array}{ll}
            \{c_1 sinh(z\sqrt{\mid a \mid}) + c_2 cosh(z\sqrt{\mid a \mid})\}\times &\\
            \{\bar{c}_1 sinh(t\sqrt{\mid b \mid}) + \bar{c}_2 cosh(t\sqrt{\mid b \mid})\}^
            {-\frac{(2 - m)}{(1 - m)}}   & \mbox { when $a<0,~~ b<0$}\\
            \{c_1 + c_2 z\} \{ \bar{c}_1 + \bar{c}_2 t\}^{-\frac{(2 - m)}{(1- m)}}   
            & \mbox { when $a = 0, ~~ b = 0$} \\
            \{c_1 sin(z\sqrt{a}) + c_2 cos(z\sqrt{a})\}\times &\\
            \{\bar{c}_1 sin(t\sqrt{b}) + \bar{c}_2 cos(t\sqrt{b})\}^{-\frac{(2 - m)}{(1 - m)}}
             & \mbox {when $a>0,~~ b>0$}
            \end{array} \right. 
\end{equation}
\noindent
With the help of Eqs. (\ref{eq22}) and (\ref{eq23}), the energy density and pressure can 
be obtained from Eqs. (\ref{eq6}) and (\ref{eq7})
\begin{equation}
\label{eq24}
32 \pi \rho = \left[ \begin{array}{ll}
              e^{-A}\{(m + 1)\mid a \mid (\frac{Z_2}{Z_1})^2 - 4 \mid a \mid + &\\
              \frac{(3-m)(2-m)^2}{(1-m)^2}\mid b \mid (\frac{T_2}{T_1})^2 - 3 \beta_0^2\}  
              & \mbox { when $a<0,~~ b<0$}\\
              e^{-A}\{\frac{c^2_2 (1 + m)}{(c_1 + c_2 z)^2} - &\\
              \frac{(3 - m)(2 - m)^2 \bar{c}_2^2}{(1 - m)^2 (\bar{c}_1 + \bar{c}_2 t)^2} 
              - 3 \beta_0^2\}  & \mbox { when $a = 0, ~~ b = 0$} \\
              e^{-A}\{4a + a(\frac{Z_3}{Z_4})^2(1 + m) + &\\
              \frac{(3-m)(2-m)^2}{(1-m)^2}b(\frac{T_3}{T_4})^2 - 3 \beta_0^2\}
              & \mbox {when $a>0,~~ b>0$}
              \end{array} \right. 
\end{equation}
\begin{equation}
\label{eq25}
32 \pi p_1 =  \left[ \begin{array}{ll}
              e^{-A}\{(m + 1)\mid a \mid (\frac{Z_2}{Z_1})^2 - \frac{4(2-m)}{(1-m)} 
              \mid b \mid + &\\
              \frac{(2-m)(2m^2 - 7m + 2)}{(1-m)^2}\mid b \mid (\frac{T_2}{T_1})^2 - 
              3 \beta_0^2\}  & \mbox { when $a<0,~~ b<0$}\\
              e^{-A}\{\frac{c^2_2 (1 + m)}{(c_1 + c_2 z)^2} - &\\
              \frac{(4 - m^2)(3 - 2m) \bar{c}_2^2}{(1 - m)^2 (\bar{c}_1 + \bar{c}_2 t)^2} 
              - 3 \beta_0^2\}  & \mbox { when $a = 0, ~~ b = 0$} \\
              e^{-A}\{(1+m)a(\frac{Z_3}{Z_4})^2 + \frac{4(2-m)b}{(1-m)} +  &\\
              \frac{(2-m)(2m^2 -7m + 2)}{(1-m)^2}b(\frac{T_3}{T_4})^2 - 3 \beta_0^2\}
              & \mbox {when $a>0,~~ b>0$}
              \end{array} \right. 
\end{equation}
\noindent
 where
$e^{-A} = cosh^{-m}(z\sqrt{\mid a \mid})cosh^{2-m}(t\sqrt{\mid b \mid}) Z_1^{-m} T_1^{2-m} \\
Z_1 = c_2 +  c_1 tanh(z\sqrt{\mid a \mid})\\
Z_2 = c_1 + c_2 tanh(z\sqrt{\mid a \mid})\\
T_1 = \bar{c}_2 + \bar{c}_1 tanh(t\sqrt{\mid b \mid})\\
T_2 = \bar{c}_1 + \bar{c}_2 tanh(t\sqrt{\mid b \mid})\\
Z_3 = c_1 -  c_2 tan(z\sqrt{a})\\
Z_4 = c_2 + c_1 tan(z\sqrt{a})\\
T_3 = \bar{c}_1 + \bar{c}_2 tan(t\sqrt{b})\\
T_4 = \bar{c}_2 + \bar{c}_1 tan(t\sqrt{b})$\\
\noindent
{\bf Case II}: Time varying displacement field $(\beta = \beta_0 t^{\alpha})$. \\   
\noindent
Using the aforesaid power law relation between time coordinate and displacement field,
Eq. (\ref{eq19}) may be written as
\begin{equation}
\label{eq26}
\ddot{w} - [\frac{3(1-m)\beta_0^2}{4(2-m)^2}t^{2\alpha} - \frac{k(1-m)}{(2-m)^2}]w = 0,
\end{equation}
\noindent
where
\begin{equation}
\label{eq27}
w = e^{-\frac{(1-m)}{(2-m)}} B_2
\end{equation}
\noindent
Now, it is difficult to find a general solution of Eq. (\ref{eq26}) and hence
we consider a particular case of physical interest. It is believed that $\beta^2$ has 
similar behaviour as the cosmological constant which decreases during expansion of
universe. Several authors {\cite{ref47}$ - $\cite{ref59}} have considered the relation 
$ \beta \sim \frac{1}{t}$ for study of cosmological models in different context.\\
\noindent
Considering $\alpha = - 1 (\beta = \frac{\beta_0}{t})$, Eq. (\ref{eq26}) reduces to
\begin{equation}
\label{eq28}
t^2 \ddot{w} + [\frac{k(1-m)}{(2-m)^2}t^2 - \frac{3}{4}\frac{(1-m)}{(2-m)^2} \beta_0^2] w = 0
\end{equation}
\noindent
Eq. (\ref{eq28}) yields the general solution
\begin{equation}
\label{eq29}
w t^{r +1} = (t^3 D)^r[\frac{c_1 e^{ht} + c_2 e^{-ht}}{t^{2r-1}}]
\end{equation}
\noindent
where \\
$ D \equiv \frac{d}{dt} \\
r = \frac{1}{2}[ \{1 + \frac{3(1-m)}{(2-m)^2} \beta_0^2\}^{\frac{1}{2}} - 1] \\
h^2 = \frac{k(1-m)}{(2-m)^2}$\\
\noindent
For $r = 1$, $\beta_0^2 = \frac{8(2-m)^2}{3(1-m)}$, Eq. (\ref{eq29}) suggests
\begin{equation}
\label{eq30}
w = (h -\frac{1}{t})c_3 e^{ht} - (h + \frac{1}{t}) c_4 e^{-ht},
\end{equation}
\noindent
where $c_3$ and $c_4$ are integrating constants.\\
\noindent
Hence the metric coefficients have the explicit forms as
\begin{equation}
\label{eq31}
e^A = \left[ \begin{array}{ll}
            \{c_1 sinh(z\sqrt{\mid a \mid}) + c_2 cosh(z\sqrt{\mid a \mid})\}^{m}\times &\\
            \{(h - \frac{1}{t})c_3 e^{ht} - (h + \frac{1}{t}) c_4 e^{-ht} \}^{(m - 2)}   
             & \mbox { when $a<0$}\\
            \{c_1 + c_2 z\}^{m} \{ (h - \frac{1}{t}) c_3 e^{ht} - (h + \frac{1}{t}) 
            c_4 e^{-ht}\}^{(m - 2)}     & \mbox { when $a = 0$} \\
            \{c_1 sin(z\sqrt{a}) + c_2 cos(z\sqrt{a})\}^{m}\times &\\
            \{(h - \frac{1}{t}) c_3 e^{ht} - (h + \frac{1}{t}) c_4 e^{-ht}\}^{(m - 1)}  
            & \mbox {when $a>0$}
            \end{array} \right. 
\end{equation}
\begin{equation}
\label{eq32}
e^B = \left[ \begin{array}{ll}
            \{c_1 sinh(z\sqrt{\mid a \mid}) + c_2 cosh(z\sqrt{\mid a \mid})\}\times &\\
            \{(h - \frac{1}{t}) c_3 e^{ht} - (h - \frac{1}{t}) c_4 e^{-ht}\}^{-\frac{(2 - m)}
            {(1 - m)}}   & \mbox { when $a<0$}\\
            \{c_1 + c_2 z\} \{ (h - \frac{1}{t}) c_3 e^{ht} - (h + \frac{1}{t}) c_4 e^{-ht}\}
            ^{-\frac{(2 - m)}{(1- m)}}   & \mbox { when $a = 0$} \\
            \{c_1 sin(z\sqrt{a}) + c_2 cos(z\sqrt{a})\}\times &\\
            \{ (h - \frac{1}{t}) c_3 e^{ht} - (h + \frac{1}{t})c_4 e^{-ht}\}^{-\frac{(2 - m)}
            {(1 - m)}}  & \mbox {when $a>0$}
            \end{array} \right. 
\end{equation}
With the help of Eq. (\ref{eq31}) and (\ref{eq32}), the energy density and pressure 
can be obtained from Eqs. (\ref{eq6}) and (\ref{eq7})
\begin{equation}
\label{eq33}
32 \pi \rho = \left[ \begin{array}{ll}
              e^{-A}\{ \mid a \mid ((\frac{Z_2}{Z_1})^2 (1 + m) - 4 ) + &\\
              \frac{(3-m)(2-m)^2}{(1-m)^2}(\frac{c_3 h^2 t}{T_6} - \frac{1}{t})^2 - 
              \frac{3 \beta_0^2}{t^2}\}  & \mbox { when $a<0$}\\
              e^{-A}\{\frac{c^2_2 (1 + m)}{(c_1 + c_2 z)^2} + &\\
              \frac{(3 - m)(2 - m)^2}{(1 - m)^2 }(\frac{c_3 h^2 t}{T_6} - \frac{1}{t})^2 
              - \frac{3 \beta_0^2}{t^2}\}  & \mbox { when $a = 0$} \\
              e^{-A}\{4a + (1 + m)a(\frac{Z_3}{Z_4})^2 + &\\
              \frac{(3-m)(2-m)^2}{(1-m)^2}(\frac{c_3 h^2 t}{T_6} - \frac{1}{t})^2  
              - \frac{3 \beta_0^2}{t^2}\}& \mbox {when $a>0$}
              \end{array} \right. 
\end{equation}
\begin{equation}
\label{eq32}
32 \pi p_1 =  \left[ \begin{array}{ll}
              e^{-A}\{(m + 1) a (\frac{Z_2}{Z_1})^2 - \frac{(1 + 2m)(2 - m)^2}
              {(1-m)^2}(\frac{c_3 h^2 t}{T_6} - \frac{1}{t})^2 - &\\
              \frac{4(2 - m)}{(1 - m)}(\frac{1}{t^2} - \frac{4 c_4 h^3 t e^{-2ht}}
              {T_6} + h^2 (\frac{T_5}{T_6})^2)- \frac{3 \beta_0^2}{t^2}\}  
              & \mbox {when $a<0$}\\
              e^{-A}\{\frac{c^2_2 (1 + m)}{(c_1 + c_2 z)^2} - \frac{(1 + 2m) (2 - m)^2}
              {(1 - m)^2}(\frac{c_3 h^2 t}{T_6} - \frac{1}{t})^2 -  &\\
              \frac{4(2 - m)}{(1 - m)}(\frac{1}{t^2} - \frac{4 c_4 h^2 t e^{-2ht}}{T_6} 
              + \frac{h^2 T_5^2}{T_6^2}) - \frac{3\beta_0^2}{t^2}\}  
              & \mbox {when $a = 0$} \\
              e^{-A}\{(1+m)a(\frac{Z_3}{Z_4})^2 - \frac{(1 + 2m)(2 - m)^2}{(1 - m)^2}
              (\frac{c_3 h^2 t}{T_6} - \frac{1}{t})^2 -  &\\
              \frac{4(2 - m)}{(1 - m)}(\frac{1}{t^2} - \frac{4 c_4 h^3 t e^{-2ht}}{T_6} 
             + h^2 (\frac {T_5}{T_6})^2) - \frac{3 \beta_0^2}{t^2}\}& \mbox {when $a>0$}
              \end{array} \right. 
\end{equation}
\noindent
where \\
$ T_5 = c_3 + c_4 (1 + 2 h t) e^{-2ht} \\
  T_6 = c_3 (ht - 1) - c_4 (1 + ht) e^{-2ht}$ \\
\noindent 
From the above results in both cases it is evident that at any instant the domain wall 
density $\rho$ and pressure $p_1$ in the perpendicular direction decreases on both 
sides of the wall away from the symmetry plane and both vanish as 
$Z \longrightarrow \pm \infty $. The space times in both cases are reflection symmetry 
with respect to the wall. All these properties are very much expected for a domain wall.\\ 
\section {Study of Geodesics}
The trajectory of the test particle $x^{i}\{t(\lambda), x(\lambda), y(\lambda), z(\lambda)\}$
in the gravitational field of domain wall can be determined by integrating the geodesic 
equations
\begin{equation}
\label{eq33}
\frac{d^2 x^{\mu}}{d \lambda^2} + \Gamma^{\mu}_{\alpha \beta} \frac{dx^{\alpha}}
{d\lambda}\frac{dx^{\beta}}{d\lambda} = 0
\end{equation}
\noindent
for the metric (5). It has been already mentioned in \cite{ref23}, the acceleration 
of the test particle in the direction perpendicular to the domain wall ( i.e. in 
the z-direction) may be expressed as
\begin{equation}
\label{eq34}
\ddot{z} = \frac{e^{B-A}}{2} \frac{\partial B}{\partial z} (\dot{x}^2 + \dot{y}^2)
- \frac{1}{2}\frac{\partial A}{\partial z}( \dot{t}^2 + \dot{z}^2) - 
\frac{\partial A}{\partial z} \dot{t} \dot{z}
\end{equation}
\noindent
By simple but lengthy calculation one can get expression for acceleration
which may be positive, negative (or zero) depending on suitable choice of the constants.
This implies that the gravitational field of domain wall may be repulsive or attractive
in nature (or no gravitational effect). \\
\section {Conclusions}
The present study deals with plane symmetric domain wall within the framework of 
Lyra geometry. The essential difference between the cosmological theories based 
on Lyra geometry and Riemannian geometry lies in the fact that the constant vector 
displacement field $\beta$ arises naturally from the concept of gauge in Lyra 
geometry whereas the cosmological constant $\lambda$ was introduced in adhoc fashion 
in the usual treatment. Currently the study of domain walls and cosmological constant 
have gained renewed interest due to their application in structure formation in the 
universe. Recently Rahaman et al \cite{ref23} have presented a cosmological model for 
domain wall in Lyra geometry under a specific condition by taking displacement 
fields $\beta$ as constant. The cosmological models based on varying displacement 
vector field $\beta$ have widely been considered in the literature in different 
context {\cite{ref38}$ - $\cite{ref42}}. Motivated by these studies it is worthwhile 
to consider domain walls with a time varying $\beta$ in Lyra geometry. In this paper 
both cases viz., constant and time varying displacement field $\beta$, are discussed 
in the context of domain walls with the framework of Lyra geometry. It has been 
pointed out that the result of Rahman et al \cite{ref23} is a special case of our 
results. Further for the sake of completeness all possible solutions of the field 
equations are presented in this paper.\\   
\section*{Acknowledgements}
\noindent
The authors would like to thank the Harish-chandra Research Institute, Allahabad, 
India for hospitality during a school on cosmology where part of this work is done. 
We also thank Shiv Sethi for useful suggestions.  
\newline
\newline

\end{document}